# Extra-dimensional confinement of quantum particles

**Eric R. Hedin**
Department of Physics and Astronomy, Ball State University, Muncie, Indiana 47306, USA

http://physicsessays.org/

Eric R. Hedin, *erhedin@bsu.edu*
*Department of Physics and Astronomy, Ball State University, Muncie, IN 47306, USA*

**Abstract.**   In this paper, a basic theoretical framework is developed in which elementary particles have a component of their wave function extending into higher spatial dimensions. This model postulates an extension of the Schrödinger equation to include a 4[th] and 5[th] spatial component. A higher-dimensional simple harmonic oscillator confining potential localizes particles into 3-d space (characterizing the "brane tension" which confines Standard Model particles to the sub-manifold). Quantum effects allow a non-zero probability for a particle's evanescent existence in the higher dimensions, and suggest an experimental test for the validity of this model via particles being temporarily excited into the first excited state of the extra-dimensional potential well, in which their probability of existing in 3-d space transiently drops to zero. Several consistency checks of the predictions and outcomes of this extra-dimensional model are included in this paper. Among the outcomes of this model are: a match with the quantum phenomenon of "*zitterbewegung*"; the predicted intrinsic spin angular momentum is of order $\hbar$; the magnetic moment of the electron is determined (with a gyromagnetic ratio of 2); the nuclear force "hard core" radius is accurately predicted; the ratio of quark masses (of the up and down quarks) is found to be $m_u / m_d = 0.56$, consistent with QCD theory; a self-consistent derivation of special relativistic effects is given; possible explanations of the Planck mass and Planck length, and a possible explanation of the origin of electric charge. In addition, a simple application of higher-dimensional particle effects to the astrophysics of stars is briefly examined, showing that radical physical inconsistencies are not evident. Finally, this model suggests a possible explanation of dark matter as the fractional probability manifestations of a ladder of the higher-dimensional symmetric excited states of ordinary particles.

**Résumé** : Cet article présente un cadre théorique dans lequel la fonction d'onde de particules élémentaires a une composante qui s'étend dans des dimensions spatiales supplémentaires. Ce modèle postule une extension de l'équation de Schrödinger et y ajoute une 4ème et une 5ème composantes spatiales. Le potentiel de confinement d'un oscillateur harmonique simple localise des particules dans l'espace à trois dimensions (caractérisant la 'tension de brane' qui confine les particules du modèle standard à la sous-variété). Les effets quantiques confèrent une probabilité non nulle à l'existence évanescente d'une particule dans les dimensions supplémentaires, et suggèrent une vérification expérimentale de la validité de ce modèle à travers des particules que l'on excite temporairement au premier état excité du puits de potentiel à dimensions supplémentaires, dans lequel leur probabilité d'existence dans un espace tridimensionnel devient transitoirement nulle. Plusieurs confrontations entre les prédictions et les résultats de ce modèle à dimensions supplémentaires sont incluses dans cet article. Au nombre de ces résultats: une correspondance avec le phénomène quantique de "*zitterbewegung*"; le moment cinétique intrinsèque prédit est d'ordre $\hbar$ ; le moment magnétique de l'électron est déterminé (avec un rapport gyromagnétique de 2); le rayon nucléaire du 'cœur dur' est prédit correctement; on trouve un rapport de masses des quarks (quarks up et quarks down) de $m_u / m_d = 0.56$, ce qui est cohérent avec la théorie QCD; une dérivation auto-cohérente d'effets relativistes restreints est incluse, de même que de possibles explications de la masse de Planck et de la longueur de Planck, ainsi qu'une explication possible de l'origine de la charge électrique. En outre, une application simple des effets de particules à dimensions supplémentaires sur l'astrophysique stellaire fait l'objet d'un bref examen, montrant que des divergences physiques radicales ne sont pas évidentes. Enfin, ce modèle avance une explication possible de la matière noire, où se manifesterait, dans des événements à probabilité partielle, une échelle des états excités symétriques de particules ordinaires dans les dimensions supplémentaires.

## 1. INTRODUCTION

The basic concept of higher spatial dimensions is not new, with a significant contribution of this line of thought in an attempt to unify electromagnetic theory and general relativity dating back to the work of Kaluza and Klein [1]. Modern string theory further extends the concept to multiple higher spatial dimensions [2, 3, 4]. An example of a significant recent work on the theory and implications of extra dimensions with finite size is given in [5]. The aforementioned theories require a "curling up" or "compactification" of the higher dimension in order to prevent unphysical manifestations. Standard Model particles and fields are said to be confined to the "brane" or submanifold, which composes our 4-dimensional spacetime (3 dimensions of space and 1 of time), but gravitational effects would be free to propagate into extra spatial dimensions unless prevented by compactification or, as Randall proposes, by a warping of the extra dimension which provides a local confinement of gravitons [4, 6, 7]. The work presented here focuses on results stemming from a generic form of a confining potential which traps Standard Model particles and fields onto the 3- (spatial) dimensional brane. It is postulated that particles are confined to 3-d space by an extra-dimensional harmonic potential well. Some of the difficulties of the earlier Kaluza-Klein theories and point-particle singularities are thereby eliminated. The size of the higher-dimensions is not constrained in this model, only the confinement scale of particles. In this model, a harmonic potential is used to confine particles and electromagnetic fields to an extremely restricted extra-dimensional region beyond the submanifold. Beyond the extent of this potential confinement, this model does not require a larger size to the extra dimensions. However, the extent of the size of the extra dimension need not be larger than that which is well below experimental constraints. A confinement potential which varies in strength with particle mass or energy, as is considered here, may be similar to the concept of a variable compactification geometry which is renormalized as a function of energy scale [8]. The consequences of allowing evanescent confinement of particles and fields in an extra-dimensional harmonic potential well include several results which provide an intuitive explanation of observed fundamental physical properties. This model also allows for testable predictions which are accessible with current technology.

## 2. RATIONALE FOR THE EVANESCENT EXTENSION OF PARTICLE WAVE-FUNCTIONS INTO HIGHER DIMENSIONS

Here, a plausibility argument is presented for the necessity of a higher-dimensional component to a particle's quantum wave function. Imagine a one-dimensional universe in which particles are confined to a line, say along the x-axis. A perpendicular to the line represents an extension into the $2^{nd}$ dimension (say along the y-axis). The x-y axes occupy a 2-dimensional (2-d) space. However, there are an infinite number of other directions into which a line perpendicular to the x-axis (but at an arbitrary angle to the y-axis) could be extended. To encompass all of these, 3-dimensional space is necessary. Similarly, if we imagine a 2-d, planar universe, (in the x-y plane), a perpendicular to it would extend into 3-d space (along the z-axis). However, by analogy, an infinite number of other directions perpendicular to the plane could be found extending into 4-d space. Thus, in general, two extra spatial dimensions are necessary (and sufficient) to encompass the infinity of perpendicular directions to a given space. Returning to the concept of a particle confined to a 1-d linear space, if by "confined" it is meant that some force or potential barrier prevents the particle from moving perpendicular to its linear universe in the ±y or ±z directions, then $\Delta y = 0$, and $\Delta z = 0$, and the Heisenberg uncertainty principle for these coordinates, $\Delta y \Delta p_y \geq \hbar/2$, and $\Delta z \Delta p_z \geq \hbar/2$, therefore indicates that the particle would have infinite momentum or energy. Extending the foregoing argument to particles in our 3-d universe implies that an evanescent extension of their wavefunctions into two higher dimensions is necessary to avoid unphysical consequences of the Heisenberg uncertainty principle for these extra-dimensional coordinates. This argument also implies that the dimensional confinement of particles cannot consist of a step function, but must rather be a "softer" dimensional confinement barrier such as the quadratic potential well used in this model.

Calculations of the constraints on the size, R, of a number n of compactified extra dimensions generally require $n \geq 2$, which allow R ≈ 1 mm [5]. However, Randall has shown that we may live in a space with n *noncompact* extra dimensions (even n=1), and still maintain compatibility with experimental gravitational measurements [6]. In this model, we have n=2, corresponding to the 4th and 5th spatial dimensions.

## 3. EXTRA-DIMENSIONAL CONFINING POTENTIAL

In considering an extra-dimensional, evanescent component to the quantum wave function of an elementary particle of mass m, our starting point is a modification of the Schrödinger equation to include 4th and 5th spatial coordinates, γ and η, which extends orthogonally out of our 3-d space into the 4th and 5th spatial dimensions. It is recognized that in this treatment we may be "sacrificing theoretical adequacy for simplicity," but as Feynmann once stated, the goal is "to see how closely our 'shadow of truth' equation gives a partial reflection of reality" [9]. Therefore, the analysis we will pursue employs non-relativistic quantum mechanics. It has been acknowledged that the standard treatment of relativistic quantum mechanics contains ambiguities which may justify seeking a deeper conceptual level of understanding [10]. Nevertheless, we will later compare results with a relativistic analysis, and see that the important features therein are reflected in our 'shadow of truth' derivation. It can be shown that the evanescent confinement of quantum particles in two extra dimensions provides a conceptual and even quantitative explanation for quantum properties of particles which are otherwise only explained within the context of relativistic quantum theory.

The Schrödinger equation with two extra dimensions is written as:

$$\frac{-\hbar^2}{2m}\left[\nabla^2 + \nabla^2_{\gamma,\eta}\right]\Psi_5(x,y,z,\gamma,\eta) + \left[U(x,y,z) + U_{\gamma,\eta}(\gamma,\eta)\right]\Psi_5(x,y,z,\gamma,\eta) = E\Psi_5(x,y,z,\gamma,\eta) \, . \tag{1}$$

$U_{\gamma,\eta}(\gamma,\eta)$ is an additive component of the potential energy of the particle which depends upon the excursion of the particle into the 4th and 5th dimensions, and which can be taken to heuristically describe the brane tension which confines Standard Model particles to the submanifold [11]. We assume that the 5-d wave function can be written in the partially separable form, $\Psi_5(x, y, z, \gamma, \eta) = \Psi(x, y, z)\Psi_{\gamma,\eta}(\gamma, \eta)$. The extra-dimensional component introduces additional degrees of freedom to the particle, which manifest in 3-d space as "intrinsic" particle properties such as spin and "*zitterbewegung.*" Inserting the 5-d wave function into Eq. (1), dividing by $\Psi(x, y, z)\Psi_{\gamma,\eta}(\gamma, \eta)$, and collecting the x,y,z-dependent terms and the $\gamma, \eta$-dependent terms separately gives:

$$\left[\frac{-\hbar^2}{2m}\frac{1}{\Psi(x, y, z)}\nabla^2\Psi(x, y, z) + U(x, y, z)\right] + \left[U_{\gamma,\eta}(\gamma, \eta) - \frac{\hbar^2}{2m}\frac{1}{\Psi_{\gamma,\eta}(\gamma, \eta)}\nabla^2_{\gamma,\eta}\Psi_{\gamma,\eta}(\gamma, \eta)\right] = E. \quad (2)$$

Each bracket on the left side of Eq. (2) must be equal to a constant. The $\gamma, \eta$-dependent terms inside the right-hand square brackets can be taken to represent a shift in the overall potential energy of the particle, as compared to its normal representation in a 3-d Schrödinger equation. It will be seen that in order to maintain consistency with the uncertainty principle, this shift will be equal to the rest-mass energy of the particle.

To determine the nature of the confining potential energy function, $U_{\gamma,\eta}(\gamma, \eta)$, we require that particles do not randomly exit from 3-d space into 4-space (where $\gamma, \eta \neq 0$), so $U_{\gamma,\eta}(\gamma, \eta)$ must have a local minimum at $\gamma = \eta = 0$. To 2$^{nd}$–order in $\gamma$ and $\eta$, this type of potential well can most generally be approximated by a symmetric quadratic harmonic oscillator form. Let $U_{\gamma,\eta}(\gamma, \eta) = k(\gamma^2 + \eta^2)/2$, which has a corresponding ground-state solution wave function, $\Psi_{0,0}(\gamma, \eta) = A\exp[-a(\gamma^2 + \eta)]$. Fig. 1 depicts this quadratic confining potential well, with the ground state probability distribution superimposed.

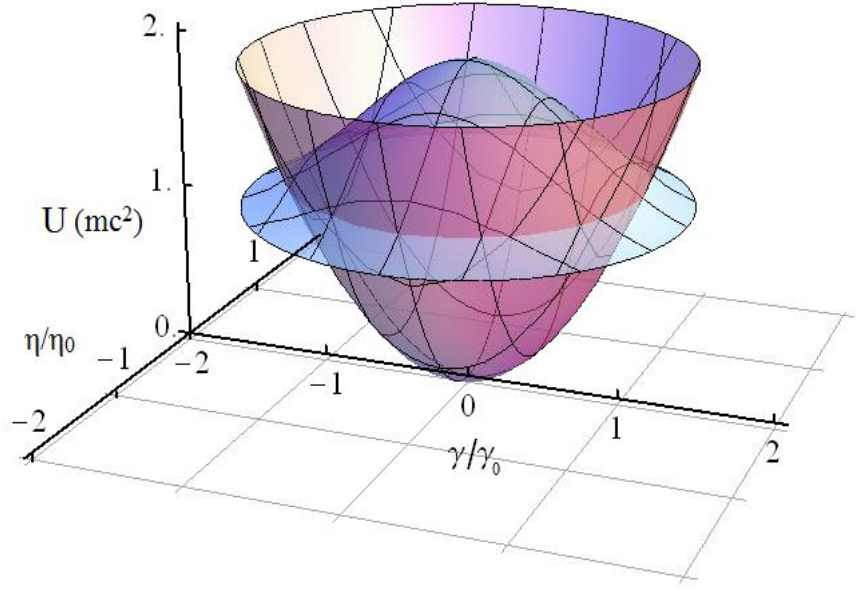

FIG. 1  (Color online)  Depiction of the quadratic confining potential well in the higher dimensions $\gamma$ and $\eta$.  The ground state probability distribution is superimposed.  3-d space intersects the higher dimensions at $\gamma=\eta=0.0$.  The $\eta$ and U-axes have been displaced for clarity.

When the above expressions for $U_{\gamma,\eta}(\gamma,\eta)$ and $\Psi_{\gamma,\eta}(\gamma,\eta)$ are plugged into the $\gamma,\eta$-dependent terms of (2), we find,

$$U_{\gamma,\eta}(\gamma,\eta) - \frac{\hbar^2}{2m} \frac{1}{\Psi_{0,0}(\gamma,\eta)} \nabla^2_{\gamma,\eta} \Psi_{0,0}(\gamma,\eta) = \frac{1}{2}k(\gamma^2+\eta^2) - \frac{\hbar^2}{2m}[4a^2(\gamma^2+\eta^2)-4a].$$

(3)

Standard analysis [12] of the harmonic oscillator shows that the right-hand side of (3) is equal to the ground state energy, $E_0$, which must be independent of $\gamma$ and $\eta$.  Thus,

$$E_0 = \frac{2\hbar^2 a}{m} + (\gamma^2+\eta^2)\left(\frac{1}{2}k - \frac{2\hbar^2 a^2}{m}\right)$$

(4)

The term in parentheses must be zero, $(k/2 - 2\hbar^2 a^2/m) = 0$, which leaves $E_0 = 2\hbar^2 a/m$. We also obtain $a^2 = mk/4\hbar^2$, so $E_0 = \hbar\sqrt{k/m} \equiv \hbar\omega_0$, with $\omega_0 \equiv \sqrt{k/m}$. The classical turning points are at the value of $\gamma$ and $\eta$ for which the ground state energy equals the potential energy: $E_0 = k(\gamma^2 + \eta^2)/2 = \hbar\sqrt{k/m}$. For this symmetric oscillator, we set $\gamma = \eta = \gamma_0$, which gives $\gamma_0^2 = \hbar/\sqrt{km} = \hbar/\omega_0 m$. We also find $\gamma_0^2 = 1/2a$, so that the ground state wave function becomes $\Psi_{0,0}(\gamma, \eta) = A\exp[-(\gamma^2 + \eta^2)/2\gamma_0^2]$, with normalization $A = 1/(\gamma_0\sqrt{\pi})$.

The Heisenberg uncertainty principle, $\Delta E\Delta t \geq \hbar/2$, is used to obtain approximate values for $\gamma_0$ and $\omega_0$ in terms of known quantities. In effect, the particle can oscillate out of 3-d space as long as it returns within the constraints of the uncertainty principle. Let $\Delta E\Delta t = \alpha\hbar/2$, where the factor, $\alpha > 1$, is to account for the extra time it takes for the particle to move out of the 3-d space due to its distributed form. If the particle moves away from $\gamma = \eta = 0$, it will disappear from our 3-d space. The time it can be gone is $\Delta t = \alpha\hbar/2\Delta E = \alpha\hbar/2mc^2$, for an isolated particle in its rest frame (below, we will discuss how under certain circumstances $\Delta E$ must be increased to take into account additional forms of energy related to the particle's motion and binding energy). We can relate $\Delta t$ to the oscillation frequency in the ground state: $\Delta t = \pi/\omega_0 = \alpha\hbar/2mc^2$; which yields, $\omega_0 = 2\pi mc^2/\alpha\hbar$. The ground state energy of the oscillator is

$$E_0 = \hbar\omega_0 = 2\pi mc^2/\alpha \tag{5}$$

By setting $\alpha = 2\pi$, the ground state energy is equated with the particle's rest-mass energy, $E_0 = mc^2$. We also then obtain the following relations, $\gamma_0 = \hbar/mc$, and $\omega_0 = mc^2/\hbar$. The expression for $\gamma_0$ is $1/2\pi$ times the Compton wavelength, $\lambda_C$ ($\lambda_C = 0.002426$ nm for electrons). The extra-dimensional Schrödinger equation (Eq. 2) is thus simply modified to include the rest mass energy of the particle:

$$\frac{-\hbar^2}{2m}\nabla^2\Psi(x, y, z) + \left[U(x, y, z) + mc^2\right]\Psi(x, y, z) = E\Psi(x, y, z) \tag{6}$$

In this model, the particle's ground state energy in the extra dimensional potential well is equal in magnitude to its rest mass energy. At the point $\gamma = \eta = 0$ (corresponding to 3-d space), the potential energy of the extra-dimensional confining potential is zero, so at this point, the ground state energy manifests as pure kinetic energy. At $\gamma = \eta = 0$, this maximum kinetic energy of the particle corresponds to a velocity of $v = \omega_0 \gamma_0 = (mc^2/\hbar)(\hbar/mc) = c$, independent of mass. This is not to be regarded as a velocity *through* 3-d space, but as oscillatory motion in and out of 3-d space into the higher dimensions. This is analogous to the motional jitter, or "*zitterbewegung*" of a particle, proposed by Schrödinger, and it exhibits approximately the same frequency and amplitude (also with a velocity c) [13]. Penrose describes *zitterbewegung* as "the electron's instantaneous motion…always measured to be the speed of light, owing to the electron's jiggling motion, even though the overall averaged motion of the electron is less than light speed." [14]

## 4. CONSIDERATIONS FROM RELATIVISTIC QUANTUM MECHANICS

Although the treatment so far concerns a free particle in 3-d space, for which the time-independent Schrödinger equation is appropriate, the confinement of the particle to the 3-d brane by the higher-dimensional harmonic confining potential may require relativistic quantum mechanics, since the ground state energy of the particle in the well is of the order of magnitude of its rest mass energy. In order to obtain a Lorentz-invariant form of the Schrödinger equation, the relativistic relationship, $p^2 c^2 + m^2 c^4 = E^2$, between energy, E, and momentum, p, provides the basis for the Klein-Gordon equation. For the case of a confining potential, V, this equation is usually written in one dimension as [15],

$$(-\hbar^2 c^2 \nabla^2 + m^2 c^4)\Psi(x) = (E - V(x))^2 \Psi(x).$$ (6)

In contrast to the above equation, Harvey [16] uses

$$(-\hbar^2 c^2 \nabla^2 + m^2 c^4)\Psi(x) = (E^2 - V^2 - 2Vmc^2)\Psi(x)$$ (7)

in his analysis of the relativistic harmonic oscillator (with $V = kx^2/2$). In the regime where the binding energy is comparable to the rest mass energy of the particle, the energy levels of the quantum oscillator are solved to show that the ground state energy, $E_0 = mc^2 + \hbar\omega/2 + 3(\hbar\omega)^2/32mc^2$. We see that the Klein-Gordon equation introduces the rest mass energy into the ground state energy of the oscillator, along with a small relativistic correction. For the case of $\hbar\omega = mc^2$, the energy difference between the first excited state and the ground state is given as $E_1 - E_0 = 11mc^2/8$, in comparison to $E_1 - E_0 = mc^2$ from the use of the Schrödinger equation in the previous section. In general, Harvey shows that relativistic corrections to the harmonic oscillator energy levels give an increase in the spacing of the levels, as compared to the non-relativistic treatment. A further interesting point manifests in Harvey's starting procedure, which employs a *potential-dependent* rest mass of the form, $m = V/c^2 + m_0$. The higher-dimensional confining potential proposed in Sec. 3, when combined with the parameters, $\omega_0 = mc^2/\hbar$ and $\gamma = \sqrt{2}\hbar/mc$ (for the turning point when η=0), can be solved for the mass to give $m = 2V/\omega^2\gamma^2 = V/c^2$, in exact agreement with Harvey.

Another treatment of the Klein-Gordon oscillator which involves an interesting parallel to the results of the higher-dimensional confining potential is given by Chargui, et al. [17]. In their solution of the one-dimensional oscillator, they include the effect of a minimal observable length (which depends inversely on the strength of the confining potential). In the higher-dimensional analysis presented here, the minimal length in which a particle can be localized is $\gamma_0 = \hbar/mc$. When this is used in their expression for the energy levels (along with $\omega_0 = mc^2/\hbar$), one obtains $E_0 = mc^2$ and $E_1 = 2mc^2$, in exact agreement with the results for the quadratic confining potential considered here.

From these two examples, it is seen that the higher-dimensional confining potential reproduces similar results to those obtained from application of the Klein-Gordon equation to the harmonic oscillator. While treating the higher-dimensional components of the quantum oscillator with various versions of the relativistic wave equation may yield more accurate refinements to the model presented in this paper, the main result of the ground-state oscillator energy, $E_0 = mc^2$, is obtained exactly from the higher-dimensional model, and the energy jump to the first excited state is nearly reproduced as well (depending on the particular analysis of the relativistic oscillator). In itself, this is a noteworthy result, that higher-dimensional considerations can empirically reproduce relativistic quantum effects.

## 5. FOURTH DIMENSIONAL EXCITED STATES

To summarize thus far, we have proposed the existence of two extra spatial dimensions into which quantum particles extend an evanescent portion of their wave function. In order to confine particles to the sub-manifold of 3-d space, it is proposed that the brane tension for Standard Model particles is in the form of a narrow, quadratic potential well in which the lowest energy state has a probability maximum centered in our 3-d space (refer again to Fig. 1).

This becomes more interesting when we consider next the first excited state of the particle in the extra-dimensional potential well. As is well known, in this state, the probability distribution function goes to zero at the origin. What this implies in our model is that in the first excited state, the particle exists solely in the higher-dimensions. Fig. 2 shows the probability distribution of a particle in one of the two degenerate first excited states of the extra-dimensional quadratic potential well. The energy level of the first excited state is $E_1 = 2mc^2$, since for this type of confining potential, the energy levels increase with quantum number, n, as $E_n = \hbar\omega_0(n_\gamma + n_\eta + 1)$, where $n = n_\gamma + n_\eta$, $n_\gamma, n_\eta = 0,1,2,...$, and $\hbar\omega_o = mc^2$. So, in order to excite a particle into the 1st excited state, it must "absorb" an energy equal to its rest mass energy. When the particle transiently jumps up to this state, a "missing energy" signature of $2mc^2$ would momentarily manifest. If there is additional energy associated with the particle, such as kinetic energy or binding energy, the $mc^2$ term must be increased accordingly, in order to maintain consistency with the uncertainty principle.

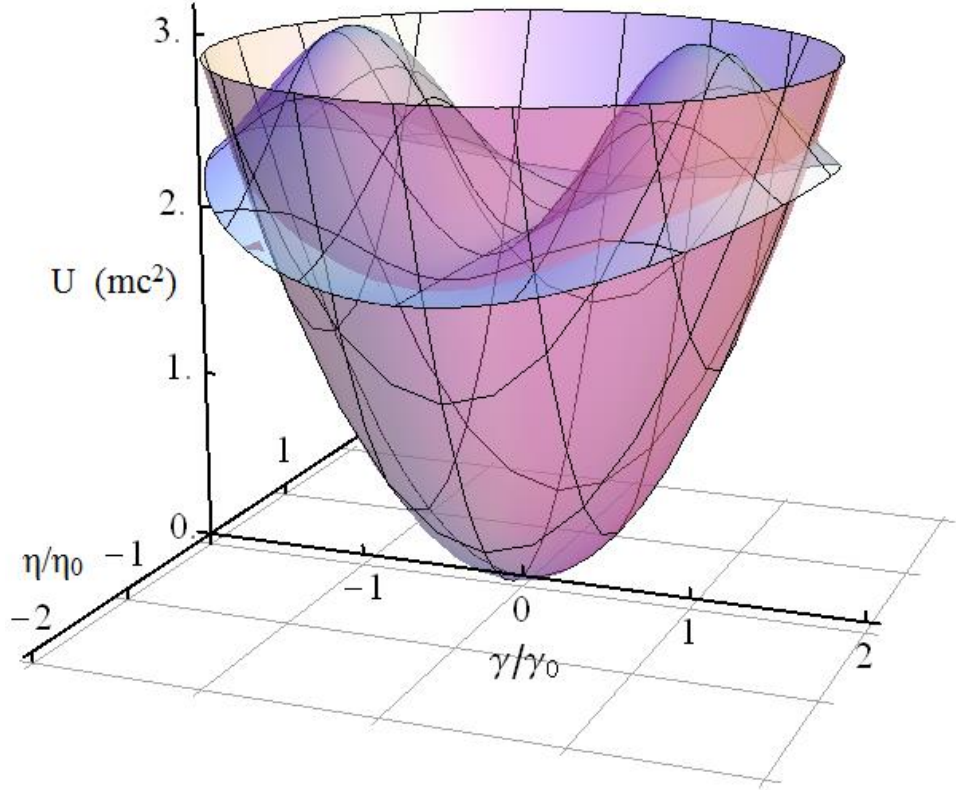

FIG. 2   (Color online)  Depiction of the quadratic confining potential well in the higher dimensions $\gamma$ and $\eta$. The excited state probability distribution (for $n_\gamma$=1, $n_\eta$=0) is superimposed, showing that in this state, the probability of finding the particle in 3-d space (at $\gamma=\eta=0.0$) drops to zero.

A Compton-scattering type experiment, in which the incident photon has energy equal to the rest-mass energy of the electron, may show the excitation effect as an absorption resonance for which the probability of detecting the scattered photon is temporarily diminished.   It may be argued that the possible excitation of particles into the higher states of the extra-dimensional confining potential would have been seen decades ago.   However, considering the electron, for example, the excitation energy would have to be specifically tuned to the electron's rest-mass energy (0.511 MeV).  A lower energy or a higher energy would not produce any higher-dimensional excitation, just as in atomic spectroscopy, where only specific photon energies are absorbed by electrons.  Now, 0.511 MeV is well above available photon energies from a tunable source—Compton Scattering experiments are usually done at X-ray energies, far below 0.511 MeV.  Particle physics experiments, on the other hand, have been conducted at much higher energies than 0.5 MeV, ever since the days of the Betatron in the early 1940's. Excitations to higher energy levels are ruled out due to the $\Delta n = 1$ selection rule discussed below. Also, the only signature of an excitation event would be a slight resonance of extremely short lifetime (estimated below to be $2 \times 10^{-19}$ sec). Other theories with compactified higher dimensions also predict "missing energy signatures" arising, however, from Kaluza-Klein graviton modes which are allowed to propagate in the higher dimensions [5, 18, 19, 20]

An estimate of how long the higher-dimensional excitation resonance lasts is found by calculating the lifetime of its first excited state. The lifetime of a stationary state would be infinite without zero-point energy to stimulate the emission. As such, the lifetime for spontaneous emission from the n[th] state is [21],

$$\tau_n = 6\pi\varepsilon_0 mc^3 / nq^2\omega^2 \ .$$ (7)

Selection rules require $\Delta n = 1$. For the oscillator frequency, $\omega$, we use $\omega = \omega_0 = mc^2 / \hbar$ to get

$$\tau_n = 6\pi\varepsilon_0 \hbar^2 / nq^2 mc \ .$$ (8)

The $1/m$ mass-dependence in Eq. (8) shows that smaller-mass particles can exist in the extra-dimensional excited state for a longer time. Comparing the excited state lifetime to the $\Delta t$ from the uncertainty principle, we obtain (using $\Delta E \Delta t = 2\pi\hbar/2$, and $\Delta E = 2mc^2$, as shown earlier),

$$\frac{\tau_n}{\Delta t} = \frac{\tau_n}{\pi\hbar / 2mc^2} = \frac{12\varepsilon_0 \hbar c}{nq^2} \ .$$ (9)

We see that $\tau_n / \Delta t$ is independent of mass, depending only upon the particle's charge, q. In terms of the fine-structure constant, $\alpha_c = q^2 / 4\pi\varepsilon_0 \hbar c$, we can write $\tau_n / \Delta t = 3/n\pi\alpha_c \approx 131$, for n=1 (transition from the 1[st] excited state to the ground state). For an electron, $\Delta t = \pi\hbar / 2mc^2 = 2.02 \times 10^{-21}$ sec, and $\tau_{n=1} \cong 2.6 \times 10^{-19}$ sec, so these times are extremely short, but there may be a measurable effect of a short-term energy loss, beyond what is masked by the uncertainty principle, if this transition to the higher – dimensional excited state is possible.

Another interesting effect would perhaps be noticeable in extremely high-temperature environments. Estimating the populations (non-degenerate) of the ground state compared to the 1$^{st}$ excited state, the Boltzmann distribution gives [22]:

$$\frac{N_1}{N_0} = \frac{e^{-E_{\gamma 1}/k_B T}}{e^{-E_{\gamma 0}/k_B T}} = e^{-(E_{\gamma 1}-E_{\gamma 0})/k_B T}.$$ (10)

The energy difference was found earlier to be $E_{\gamma 1} - E_{\gamma 0} = mc^2$, so

$$N_1/N_0 = e^{-mc^2/k_B T}.$$ (11)

At "room temperature", $k_B T = 0.2525$ eV, for $T = 293$ kelvin. For an electron, with rest mass energy, $mc^2 = 0.511x10^6$ eV, we obtain $N_1/N_0 = e^{-2x10^7} \approx 0$, as one would expect (under normal conditions, particles do not just disappear into higher dimensions!). But if $k_B T = mc^2$, we find that $N_1/N_0 = e^{-1} \approx 0.368$, implying that a significant fraction of the particles would exist in the 1$^{st}$ excited state, and thus appear to be missing from our 3-dim. space (Coulomb binding energy contributions for the electrons drastically reduce this fraction, as discussed in the section below on astrophysics). For electrons, $k_B T = mc^2$ implies a temperature of 6x10$^9$ Kelvin, a temperature on the order of what would exist in the core of a massive star at the last stage of its fusion cycles when its iron core collapses, resulting in a supernova explosion [23]. The loss of electrons into the higher dimensional excited state would cause a drop in pressure in the core of the star, perhaps contributing to its collapse. Furthermore, since the heavier protons ( $m_p >> m_e$ ) would have a much smaller fraction of their number in the higher-dimensional excited state, it could happen that a charge imbalance develops in the core (unless electrical charge somehow "leaks through" from the extra dimensional excited state into our 3-d space). This charge imbalance would also destabilize the core, perhaps contributing to its explosive end. This would occur as the protons, which are left in disproportionate numbers in 3-d space, repel each other radially outwards (source of high-energy cosmic rays?). The electrons which have moved into the higher-dimension excited state would perhaps be left behind as the core collapses into a black hole, giving the black hole a net negative charge. This topic is explored further in a later section.

Even higher temperatures are postulated in the early phases of the big bang in the early universe. The possibility of even heavier particles existing in excited states of the extra-dimensional potential may have had effects on the details of the expansion of the early universe.

## 6. PHOTONS

In this Section, some conceptual speculations are offered as to the nature of photons in relation to the idea of higher-dimensional confinement. As seen earlier, the "*zitterbewegung*" speed of the particle as it oscillates in and out of our 3-d space is always the speed of light, c. Within this model, then, it is suggestive that a photon is a "particle" which has had its oscillatory motion rotated so that this "*zitterbewegung*" motion is parallel to our 3-d space, instead of perpendicular to it. Conservation of momentum and conservation of specific quantum numbers would prevent this rotation from happening under normal circumstances. Since, in the frame of reference of a photon, the universe is Lorentz-contracted to zero size, it may still experience confinement in a higher-dimensional potential well encompassing our entire universe. From the frame of reference of any observer in the 3-d universe, however, the effective $\gamma_0$ of the photon is infinite (or at least the scale length of the universe), which according to this model and the definition of $\gamma_0$ given in Sec. 3, explains the zero mass of the photon: $m_{photon} = \hbar/\gamma_0 c \rightarrow 0$, for $\gamma_0 \rightarrow \infty$.

The photon's oscillations within this well can be considered to manifest in our 3-d space as photons moving forward through space and time, and (on the "back-swing") as anti-photons moving backwards through space and time. Photons are their own anti-particles, and an anti-particle can be considered as a particle moving backwards through time [24]. We only "see" photons, and not anti-photons, since our perception of time is "rectified"—always moving forward in time. However, under certain conditions, the effects of an anti-photon's backward movement through space and time can be observed. One example of this is with entangled photons; when one of the pair is measured in some way, the other "knows" instantly the outcome of that measurement, and modifies its behavior accordingly [25]. This has presented a mystery of apparent "spooky action at a distance" [26], or more colorfully, "passion at a distance" [27] but with this model, it is communication through time, not space, which occurs between the photons.

Due to its oscillatory nature in the higher-dimensional potential well, the photon is continuously moving between its current position in space and time (as measured in the lab) and its starting position in space and time. It begins this process the moment it is emitted, and stops the moment it is absorbed. The lab observers only see the normal "forward" motion, and to them, the photon is always where one would expect it at each moment of their time. One can say that a photon is a superposition of a quantum state moving forward in space and time, and one moving backwards in space and time. After a measurement has occurred on a photon, which, say, determines its polarization state, from that moment on (both forwards and backwards in time) the photon exists in that particular polarization state. The anti-photon, moving backwards in time, carries this particular polarization state back to the origin, where it then determines the polarization state of its entangled partner photon. When this photon is later measured, it is found to be in the appropriate, complementary polarization state (no surprise!). Entangled photons can also be described as "one particle" from the point of view of the photons, which by traveling at the speed of light, "see" a universe of zero size in their direction of motion. Thus they can always affect each other, since they are never separated. I believe that this description is consistent with the higher-dimensional model given above.

## 7. FURTHER CONSIDERATIONS AND IMPLICATIONS

### A. Dimensional confinement

Using the parameters derived earlier, we can estimate the force necessary to move a particle out of 3-d space into the 4th or 5th dimension. For the proposed quadratic potential, the force can be given by $F_\gamma = -m\omega_0^2\gamma = -\left(m^3c^4/\hbar^2\right)\gamma$. To move the particle to the turning point at $\gamma_0 = \hbar/mc$, the required force is $m^2c^3/\hbar$. For an electron, this amounts to 0.212 N, which is enormous when considering only a single electron. So the idea of forcibly moving a macroscopic object into the $4^{th}$ dimension is prohibited.

## B. Spatial curvature

It is proposed that the oscillatory motion of particles in the higher-dimensional potential well can be alternatively viewed as a localized bending of 3-d space back and forth into the higher dimensions. The nominal radius of spatial curvature localized on a given particle is given by $\gamma_0 = \hbar/mc$: a particle of small mass only shallowly "indents" 3-d space; a heavier particle more sharply bends space. Another theoretical treatment which produces a similar concept is the non-relativistic limit of the Dirac equation, in which a term known as the "Darwin term" appears [28]. It is interpreted to mean that in this relativistic theory the electron's position is diffused over a volume of order $(\hbar/mc)^3$, as a consequence of "*zitterbewegung*." From General Relativity, the magnitude of the dimensionless quantity, $Gm/rc^2$, can be regarded as the curvature of space or the strength of the gravitational field at r, due to mass, m, within a sphere of radius r [29]. In particular, if $Gm/rc^2 = 1/2$, the mass forms a black hole with event horizon at radius r. If we set $r = \gamma_0 = \hbar/mc$, we obtain, $Gm/rc^2 = Gm^2/\hbar c$, which can be considered a measure of the quantum curvature for an elementary particle of mass m whose "radius" is nominally defined by the magnitude of its oscillatory excursions into higher-dimensional space. This factor, $Gm^2/\hbar c$, is far less than $1/2$ for known elementary particles; for example $Gm^2/\hbar c = 1.75 x 10^{-45}$ for an electron, indicating that such particles are not to be considered as miniature black holes. However, an interesting question is: At what mass is the "quantum curvature", $Gm^2/\hbar c = 1/2$? We find $m = \sqrt{\hbar c/2G} = 1.54 x 10^{-8}$ kg, which is $m_{pl}/\sqrt{4\pi}$, where $m_{pl} \equiv \sqrt{\hbar c/G}$ is the Planck mass [30]. This result can be interpreted to mean that the smallest possible black hole has a mass of $m_{pl}/\sqrt{4\pi}$; at any smaller mass than this, higher-dimensional quantum effects keep the spacetime curvature sufficiently smoothed out so that a black hole cannot form. One could also say that $m_{pl}/\sqrt{4\pi}$ is the largest possible mass of an elementary particle; any particle with a larger mass would become a black hole. Other authors have previously suggested that an elementary particle may behave like a black hole if its mass is equal to the Planck mass so that its Compton wavelength, $\lambda_C = h/mc$, is therefore approximately equal to its Schwarzchild radius [31]. The model presented here provides a conceptual framework in support of this result. The Planck radius is related to $\gamma_0 = \hbar/mc$ when $m$ is the minimum mass for a black hole:

$$\gamma_0 = \hbar/(m_{pl}/\sqrt{4\pi})c = \sqrt{2\hbar G/c^3} = r_{pl}/\sqrt{\pi} = 2.3x10^{-35} \text{m}.$$ The Planck radius can then be interpreted as approximately the amplitude of the higher-dimensional oscillations of a particle whose mass is the minimum mass for a black hole.

With the interpretation of $\gamma_0 = \hbar/mc$ as the radius of curvature of 3-d space due to a particle's oscillations in the higher-dimensional potential well, we also conclude that $\gamma_0$ is a measure of the minimum size (radius) of an elementary particle of mass m. Relativistic effects would modify $m \rightarrow m/\sqrt{1-v^2/c^2}$, effectively increasing the mass to account for the kinetic energy of the particle, which in turn implies a smaller effective size. For example, a 200 GeV electron would have an effective size of about $10^{-18}$ m. This effect exactly parallels the implications of *zitterbewegung* which result from analysis of the free particle solution of the Dirac Equation in the non-relativistic limit, as mentioned above [32]. Relativistic effects manifest as the particle's velocity approaches the speed of light and produce a decrease in the amplitude of the *zitterbewegung* oscillations, which consequently causes a smaller effective size of the particle. The results of the higher-dimensional theory presented here are entirely consistent with these effects.

Another interesting corroboration of this concept comes from an estimate of the range of the repulsive core of the strong nuclear force, which according to this model would be given by $\gamma_0 = \hbar/mc = 0.21$ fm, using m as the mass of the proton. This value matches well the estimate of 0.2 fm found from experiment [33].

## C. Intrinsic spin

The properties of higher-dimensional geometry are not easily visualized, but some of its unique features can be understood by comparing our 3-d space to a 2-d planar world, and then extrapolating the comparison up to higher dimensions. Higher-dimensional space encompasses our 3-dimensional space. This concept is also implied in recent work on higher-dimensional space-times in which standard model particles are localized on a 3-brane in higher-dimensional space [5, 9]. "In 4-d geometry…a line can pass through a point of the interior of a solid without passing through any other of its points." [34] This last statement relates to how a quantum particle could "oscillate" along a line in and out of the 4$^{th}$ or 5$^{th}$ dimensions through a single point of our 3-d space. In the previous section, we saw how the confinement parameter, $\gamma_0 = \hbar/mc$ , could be considered as a radius of spatial curvature around a particle of mass m. We now consider that the particle's direction of oscillation, although always perpendicular to our 3-d space, may in fact oscillate along an infinite number of lines which intersect the position of the particle from different directions extending into the two extra dimensions (see Fig. 3).

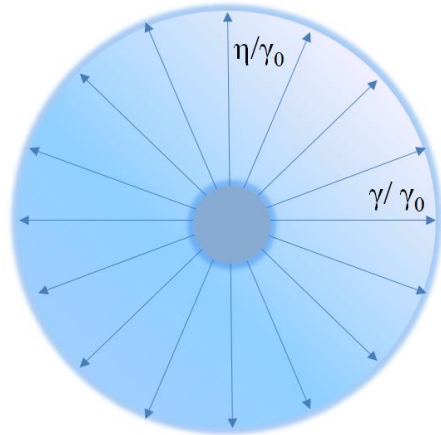

FIG. 3 (Color online) View of the "circularly polarized" extra-dimensional oscillations of a quantum particle along lines perpendicular to 3-d space.

This effect can be compared to polarized light, which although it may be polarized along, say, the x-axis, still has components of the oscillation of its electric field vector at other angles. By shifting the phase of one vector component of linearly polarized light, it may be transformed into circularly polarized light. For the purpose of conceptualizing the intrinsic spin of an elementary particle, we assume that its oscillation in and out of the 4th and 5th dimensions is "circularly polarized." The particle has an angular frequency of oscillation, $\omega_0 = mc^2/\hbar$. The resulting "motion" would manifest in our 3-d space as a particle of nominal radius, $\gamma_0 = \hbar/mc$, which has an intrinsic spin angular momentum of $L = I\omega_0 = (km\gamma_0^2)\omega_0 = km(\hbar/mc)^2(mc^2/\hbar) = k\hbar$. $I = km\gamma_0^2$ is the moment of inertia of the particle, where the factor $k$ describes the effective mass distribution of the particle and must be less than one if the mass has a radial extent of $\gamma_0$. To match the intrinsic spin angular momentum of fermions (the measurable component along a given axis), we can set $k = 1/2$.

The interpretation outlined above shows that the spin angular momentum of particles as predicted by this model is independent of mass and is of order $\hbar$, matching the quantum mechanical features of fermion spin, derived from relativistic quantum mechanics. Also, the concept of the spin arising from the rotating oscillation direction of a particle along higher-dimensional lines piercing our 3-d space, is consistent with the quantum mechanical notion of the spin as an *intrinsic* property of the particle not associated with the simple rotation of a mass. Furthermore, this spin, as outlined above, would not have any preferred direction in 3-d space (originating from motion in and out of our space into the higher dimensions), apart from a quantized component along a given measurement axis.

An alternative way of calculating the electron spin angular momentum which avoids having to set the free parameter, $k = 1/2$ is outlined below. It is proposed that the mass density of a particle is proportional to its probability distribution, where the square of the evanescent higher-dimensional wave function of the particle also represents the intrinsic probability distribution of the particle in 3-d space, due to its oscillation in the higher-dimensional potential well which results in a local curvature of 3-d space. The intrinsic mass density of the particle is then taken to be, $\rho(r) = \rho_0 \exp(-r^2/\gamma_0^2)$. The mass of the particle is found by a normalization procedure in which this density is integrated over all space (assuming a free particle).

$$m = 4\pi\rho_0 \int_0^\infty r^2 e^{-r^2/\gamma_0^2} dr = 4\pi \frac{\sqrt{\pi}}{4} \rho_0 \gamma_0^3 . \qquad (12)$$

From Eq. (12), $\rho_0$ can be expressed in terms of the particle's mass. Next, the increment of mass is found as $dm = \rho dV = \rho 4\pi r^2 dr$. This is used in determining the particle's intrinsic moment of inertia,

$$I = \int_0^\infty r^2 dm = \frac{4m}{\gamma_0^3 \sqrt{\pi}} \int_0^\infty r^4 e^{-r^2/\gamma_0^2} dr = \frac{3}{2} m \gamma_0^2 . \qquad (13)$$

The particle's total intrinsic spin angular momentum as a manifestation of its oscillation in the higher-dimensional potential well is then, $L = I\omega_0 = (3/2)\hbar$. This angular momentum is not about any specific rotation axis in 3-d space since the "rotation" originates from the oscillation in the higher-dimensional potential well, and is spread uniformly along every axis through the particle's location in 3-d space. Measured along any given axis (x, y, or z), the component of angular momentum would be 1/3 of the total, analogously to kinetic theory in which the average kinetic energy per degree of freedom of a particle is $k_B T/2$, where for 3-dimensions of freedom, the total average kinetic energy is $3kT/2$. For example, the angular momentum measured about the z-axis would be $L = (1/2)\hbar$   This calculated value is of course in agreement with the nominal value of $\hbar/2$ from standard quantum theory. It may be noted that all experimental measurements of fermion spin actually measure the fermion's magnetic moment, and then assume a certain relationship between the magnetic moment and the spin angular momentum.

**D.  Electric charge**

In considering the nature of electric charge, all that is really known is that a charged particle acts as a nearly point-source of electric (and magnetic) field. With this model, it is suggested that the electric field of a particle such as an electron and its corresponding magnetic dipole moment are the 3-dimensional cross-sections of an electromagnetic standing wave in the higher-dimensional potential well. If this is so, the ratio of the electric to the magnetic field strength at the nominal particle radius $r = \gamma_0 = \hbar/mc$, should be $E/B = c$, as is the case for all electromagnetic waves. The magnetic field at distance r along the axis of a magnetic moment $\mu_z$ is $B_z(r) = \mu_z/2\pi\varepsilon_0 c^2 r^3$. The electric field at a distance r from the point charge, e, is $E = e/4\pi\varepsilon_0 r^2$. The ratio is $E/B = ec^2 r/2\mu_z$. Plugging in the nominal particle radius for an electron, $r = \gamma_0 = \hbar/m_e c$, we obtain $E/B = ec\hbar/2m_e\mu_z$, which becomes $E/B = c$ if the electron magnetic moment has its accepted value of $\mu_z = e\hbar/2m_e$. Note that this method yields the correct gyromagnetic ratio (g=2) for the electron, which in standard quantum theory is a consequence of the relativistic Dirac equation [35]. In order for the electric field and the magnetic moment to manifest in 3-d space with no preferential angular dependence, we assume that the electromagnetic standing wave exhibits the higher-dimensional analog of circular polarization.

## E. Magnetic moment

If the mass and charge in a rotating object have the same distribution, it can be shown that the magnetic moment due to the rotation of the charged object is $\mu = (Q/2m)I\omega$, where Q is the total charge, m is the mass, $I$ is the object's moment of inertia about the rotational axis, and $\omega$ is its angular velocity. From the section on intrinsic spin, we have $I = m\gamma_0^2/2$. Plugging in for $\gamma_0$ and $\omega = \omega_0 = mc^2/\hbar$, the magnetic moment for an object of elementary charge, e, is then $\mu_z = (1/2)e\hbar/2m$. This result, when applied to the electron gives $\mu_B/2$ ($\mu_B = e\hbar/2m_e$ is the Bohr magneton), which is $1/2$ the accepted value, indicating that the charge of the electron is not distributed the same as the electron mass. The result from the section above, however, gives a prediction of the electron magnetic moment which agrees with the accepted value, indicating perhaps a preferable conceptual model of the origin of the electric field and magnetic moment of the particle.

## F. Quarks

The magnetic moments of the proton and neutron are complicated by the internal quark makeup of these baryonic particles. However, using the conceptual model of the electric charge and magnetic moment as the 3-d cross-section of a higher-dimensional electromagnetic standing wave, one can arrive at reasonable predictions involving the magnetic moments of the proton and the neutron in terms of the up and down quarks. In connection with the magnetic moment calculation, this model predicts masses of the up and down quarks.

As stated in the section above on electric charge, at the nominal radius of $\gamma_0 = \hbar/mc$, it is postulated that the ratio of the electric field and the magnetic field due to the magnetic moment of the particle should be that of an electromagnetic wave: $E/B = c$. For a particle of charge q and mass m, we find, as shown above, $E/B = qc\hbar/2m\mu_z$. Setting this ratio equal to the speed of light, c, yields $1 = q\hbar/2m\mu_z$, or $\mu_z = q\hbar/2m$ for the particle's magnetic moment aligned with a given axis. To apply this with the up and down quarks which make up the proton and neutron, we define, $\mu_u = 2e\hbar/6m_u$, and $\mu_d = -e\hbar/6m_d$, where $m_u$ and $m_d$ are the masses of the up and down quarks, respectively. The electric charges of the up and down quarks are the standard values of $+2e/3$ and $-e/3$, respectively. We now suppose that the proton and neutron magnetic moments are composed of the addition of the appropriate combination of quark magnetic moments. The negative and positive quarks are assumed to have opposite polarization rotation directions so that their contributions to the overall magnetic moment of the proton or neutron are cumulative. Using the measured values of the proton and neutron magnetic moments [36], one finds the following ( $\mu_N \equiv e\hbar/2m_p$ ):

$$\mu_p = 2.79285\mu_N = 2\mu_u - \mu_d \quad \text{and} \quad \mu_n = -1.91304\mu_N = -\mu_u + 2\mu_d . \tag{14}$$

These two equations contain the two quark masses as unknowns, and their solution gives,

$$m_u = 0.54456m_p \quad \text{and} \quad m_d = 0.96784m_p . \tag{15}$$

When the quarks are combined to form hadrons, their mutual binding energy results in a combined mass which is less than their constituent masses. Quark masses are not independently measured and are model-dependent. However, certain models based on the chiral perturbation theory of QCD give values for the ratio of the masses of the up and down quarks [37]. This ratio is given as $m_u/m_d = 0.56$, which matches the ratio of quark masses predicted by this model: $m_u/m_d = (0.54456/0.96784) = 0.56265$. This agreement again suggests the validity of the higher-dimensional electromagnetic wave model for the electric charge and magnetic moment of an elementary particle.

## G. Relativistic motion

We now analyze the motion of a particle through 3-d space, considering its extra-dimensional oscillation perpendicular to 3-d space. However, the sense of "perpendicular" is relative, depending upon the relative velocity between the observer and the particle. In Fig. 4, we show three sketches of a particle moving at different velocities through 3-d space relative to the speed of light, c (v=0, at rest in the frame of reference of the "lab" observer; v<c in the "lab" frame; and v≈c in the "lab" frame). The particle's oscillatory motion (one-half an oscillation period) into and out of the higher dimension is also shown in each frame. The higher-dimensional oscillatory displacement in the frame of reference of the lab observer is $\gamma_0$; and $\gamma'$ is how $\gamma_0$ transforms in the particle's frame of reference, according to the lab observer when v≠0.

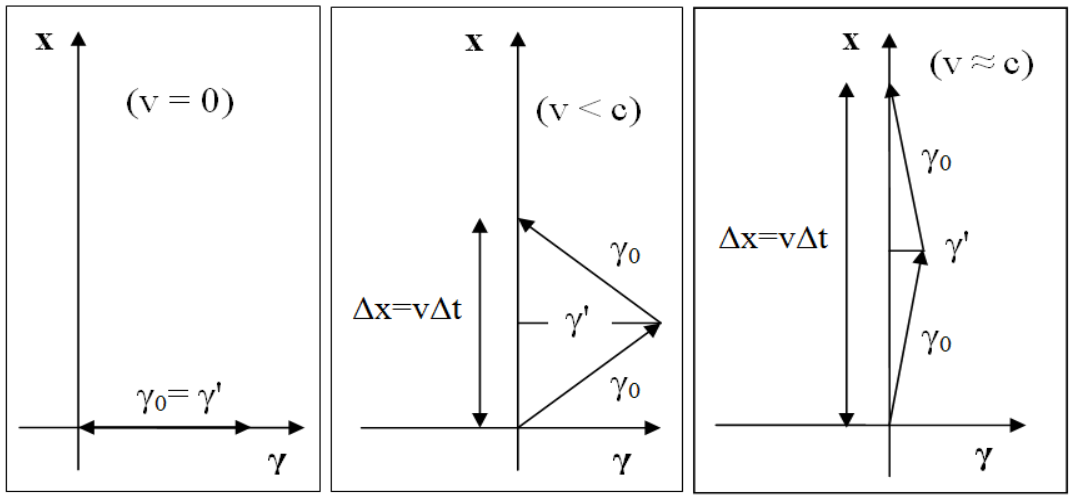

FIG. 4   One-half period of the higher-dimensional oscillation of a particle moving through 3-dimensional space at three different velocities (v=0, v<c, and v≈c) as seen the frame of reference of the "lab" observer.

The oscillatory motion of the particle is discussed by Penrose in conjunction with an electron as "zig" and "zag" particles which are continuously converted into one another as the electron moves through space [38]. However, Penrose does not interpret the oscillatory motion as extra-dimensional and makes no connection from it to relativistic effects, as is done below. Inspecting the diagram in Fig. 4, one sees the relation, $\left(\Delta x/2\right)^2 + \gamma'^2 = \gamma_0^2$. We next replace the displacement variables with the appropriate time variables, as follows: $\Delta x = v\Delta t$, $2\gamma' = c\Delta t_0$, and $2\gamma_0 = c\Delta t$, where from the discussion of *zitterbewegung* in Sec. 3, the oscillation speed of the particle in and out of higher-dimensional space is c. The oscillation distance $\gamma_0$ is what the particle "experiences" in its own (proper) frame of reference. It is associated with a turn-around time of $\Delta t_0$, which is the time an observer in the particle's proper frame of reference would use in conjunction with the Heisenberg uncertainty principle for calculating the maximum time which that particle could exist in higher-dimensional space without violating conservation of energy (see Sec. 3). This concept provides a natural explanation for the speed of light, c, being a maximum speed of any particle—if the particle moves faster than c, the higher-dimensional oscillation cannot "keep up" with the particle. Inserting time variables into the Pythagorean relation in place of the displacement variables, we obtain, $\left(v\Delta t/2\right)^2 + \left(c\Delta t_0/2\right)^2 = \left(c\Delta t/2\right)^2$. Upon rearrangement, this becomes $\Delta t = \Delta t_0 \Big/ \sqrt{1-v^2/c^2}$, in which the particle's frame of reference (proper time) is expressed by $\Delta t_0$ and "lab time" is expressed by $\Delta t$. The relative speed between the particle and the lab observer is v. This is of course the time dilation formula of Einstein's special relativity theory. In the context of this model, it can be seen how this result also follows from the postulate of extra-dimensional confinement.

We now return to the Pythagorean relation, which indicates that the effective oscillation length of the particle contracts from $\gamma_0$ to $\gamma'$. The relation is $\gamma'^2 = \gamma_0^2 - \left(\text{v}\Delta t/2\right)^2$, and we use $2\gamma_0 = c\Delta t$, from above, to eliminate $\Delta t$, obtaining, $\gamma' = \gamma_0\sqrt{1 - \text{v}^2/c^2}$. The nominal definition of the higher-dimensional oscillation length is $\gamma_0 = \hbar/mc$, so the contraction of $\gamma_0$ to $\gamma'$ indicates an effective increase in the mass of the moving particle over its rest mass, m: $m' = m\big/\sqrt{1 - \text{v}^2/c^2}$. This result is also consistent with relativistic dynamics in which the total energy of a moving particle increases over its rest mass energy as: $E = mc^2\big/\sqrt{1 - \text{v}^2/c^2}$. These results of course imply the correct relativistic energy-momentum relation, $E^2 = p^2c^2 + m^2c^4$. In relativistic quantum theory, this relation is the starting point for the development of the Klein-Gordon and Dirac Equations. However, in this model, it follows as a consequence of the assumption of a higher-dimensional confining potential.

## H.  Astrophysics

Under normal conditions, the extra-dimensional nature of particles does not manifest in any unnatural way. But under extreme conditions, this theory predicts measurable manifestations related to particle transitions into the 1st-excited state of the higher-dimensional potential well. The possibility of energy from supernovae being partially radiated into higher-dimensional channels by means of Kaluza-Klein gravitions has been considered in other works, with consequent upper bounds on the radius of the compactified dimensions [10, 39]. Some astrophysical consequences of the predictions of this model in the regime of the high temperatures in the cores of stars are considered below.

As noted earlier, the energy level difference between the first excited state and the ground state for a particle of mass m in the higher-dimensional potential well is $mc^2$. For the case of non-degenerate stellar cores, Maxwellian statistics applies (degeneracy will be considered later). Shown earlier, in Eq. (11), the fraction of electrons in the 1st-excited state (where they have zero probability of existing in 3-d space) is $\exp(-mc^2/k_BT)$. However, it must be remembered that in the exponent, $mc^2$ was taken as the total energy of the particle, which applies when it is isolated and at rest with respect to a "lab frame." For moving particles, the total energy must include the kinetic, or thermal energy: $E = KE + mc^2$. In the core of stars (non-relativistic), $KE = 3k_BT/2$, so the excited fraction becomes $\exp[-(3k_BT/2 + mc^2)/k_BT] = \exp(-3/2)\exp(-mc^2/k_BT)$, which gives a small correction. Another factor which must be considered is the *change* in the Coulomb self-energy of the star if a charge disappears into the higher dimensions, which is the same energy it would take to remove the electron from $r_0$ (its radial position from the center of the star) to $r \rightarrow \infty$. This Coulomb energy term is $E_C = kQe/r_0$, where $k = 1/4\pi\varepsilon_0$, and $Q = Ze$ is the net positive charge due to $Z$ protons inside a sphere of radius $r_0$, measured from the center of the star. To remove a single electron gives only a negligible contribution to $E_C$ ($\approx 3.6 x 10^{-16}$ eV). However, this effect will build up substantially after about 200 Coul. of charge is excited from the core into the higher-dimensions. The Coulomb self-energy becomes significant when $E_C \approx m_e c^2 \approx 0.511\,\text{MeV}$. We consider an average stellar core radius of $r = 4 x 10^6\,\text{m}$, for which $E_C = kZe^2/r = 0.511\,\text{MeV}$ yields $Z = 1.4 x 10^{21}$ electrons, corresponding to about 223 Coul. of charge. (The radius of the layered core, out to the hydrogen-burning shell, in a post main-sequence star is about $7 x 10^6\,\text{m}$.)

Another point to consider concerning net charge in a star is the balance between the outward electric force due to Coulomb repulsion and the inward gravitational attractive force. This balance is given by [40]

$$\frac{kZ_{net}e^2}{r^2} \leq \frac{GMm}{r^2} < \frac{GAm_p^2}{r^2},\qquad(16)$$

where $Z_{net}e$ is the net charge on the star inside a radius, $r$. $M$ is the star mass, $m_p$ and $e$ are the mass of the proton and its charge, and $A$ is the number of baryons in the star, inside of a radius $r$. Inserting numerical values for the physical parameters yields $Z_{net} < 10^{-36}A$. For a solar mass core $\left(M_{Sun} = 2x10^{30} \text{ kg}\right)$, the number of baryons is $A \approx 1.2x10^{57}$, so the maximum excess number of charges on a star is $Z_{net} < (10^{-36})(1.2x10^{57}) = 1.2x10^{21}$, which is nearly the same as obtained above when considering the net charge in the core which would give a Coulomb energy comparable to the electron rest mass.

So, a possible scenario is as follows:  A star core gravitationally compresses and heats up, exciting a certain fraction of electrons into the 1$^{st}$ excited state of the higher-dimensional potential well.  This leaves a net charge, consisting of positive ions, in the core of the star, which are forced radially outward against gravity if $Z_{net} > 10^{-36}A$, where both the net charge, $Ze$, and the baryon mass, $Am_p$, are symmetrically distributed inside the spherical core of a given radius.  If $Z_{net} < 10^{-36}A$, the force of gravity holds the net charges inside that radius.  As the temperature in the core increases further, at later stages of fusion burning, the fraction of electrons moving into the 1$^{st}$ excited state increases and would cause the net charge to greatly exceed the limit ($Z_{net} < 10^{-36}A$), causing an explosive build-up of Coulomb self-energy ($E_{self} = 3k(Ze)^2/5r$, for the net charge, $Ze$, evenly distributed throughout the spherical volume of radius $r$).  However, this scenario is well-moderated by the Coulomb energy term ($E_C = kQe/r_0$) in the exponent of the Maxwell-Boltzmann distribution which governs the fraction of electrons excited into the 1$^{st}$ excited state of the higher-dimensional potential well: $N_1/N_0 = \exp(-3/2)\exp[-(mc^2 + E_C)/k_BT]$. The Coulomb energy term becomes significant in its restraining effect on higher-dimensional excitation of electrons at just the conditions which would otherwise lead to runaway excess net charge build-up in the hot stellar core.

An interesting point is that only a *minute* fractional charge imbalance in a solar mass core would give a self-energy ($E_{self} = 3k(Ze)^2/5r$) equivalent to the total annihilation energy of the core mass.  For a solar-mass core of radius $r = 4x10^6$ m, one obtains, $3k(Ze)^2/5r = M_{Sun}c^2 \approx 2x10^{47}$J, yielding $Ze \approx 1.2x10^{22}$ Coul., corresponding to about $7x10^{40}$ elementary charges.  Compared to the approximately $6 \times 10^{56}$ protons in one solar mass, this fractional charge is $1 \times 10^{-16}$. Even a slight charge imbalance in a star represents a prodigious amount of electrical energy.

We next consider at what temperature the fractional charge in the core reaches the critical value of $Z_{net}/A = 10^{-36}$, at which the outward electric force equals the inward gravitational force on the excess charges. For post main-sequence stellar cores, the total number of protons is ½ the total number of nucleons, A. We therefore set, $\exp(-3/2)\exp[-(mc^2 + E_C)/k_B T]/2 = 10^{-36}$ to obtain T=1.4x10$^8$ K. This temperature is reached at the end of the H-burning phase (T~10$^8$ K), or during the He-burning phase (T~1 to 7.5x10$^8$ K) of a large star [41]. It is just at the end of the H-burning phase of a star that the outer envelope of the star is greatly expanded. As the temperature of the core continues to increase with subsequent fusion stages, the fraction of electrons in the higher-dimensional 1st excited state continues to rise, causing with each temperature increase an outflow of positive ions when the fractional charge exceeds $Z_{net}/A \approx 10^{-36}$.

The dependence of the fractional charge, $Z_{net}e$, on the core temperature can be estimated by solving the equation,

$$Z_{net}/A = \exp(-3/2)\exp[-(mc^2 + kZ_{net}e^2/r)/k_B T]/2 \qquad (17)$$

for $Z_{net}$. This transcendental equation has been solved numerically, with the results presented graphically in Fig. 5.

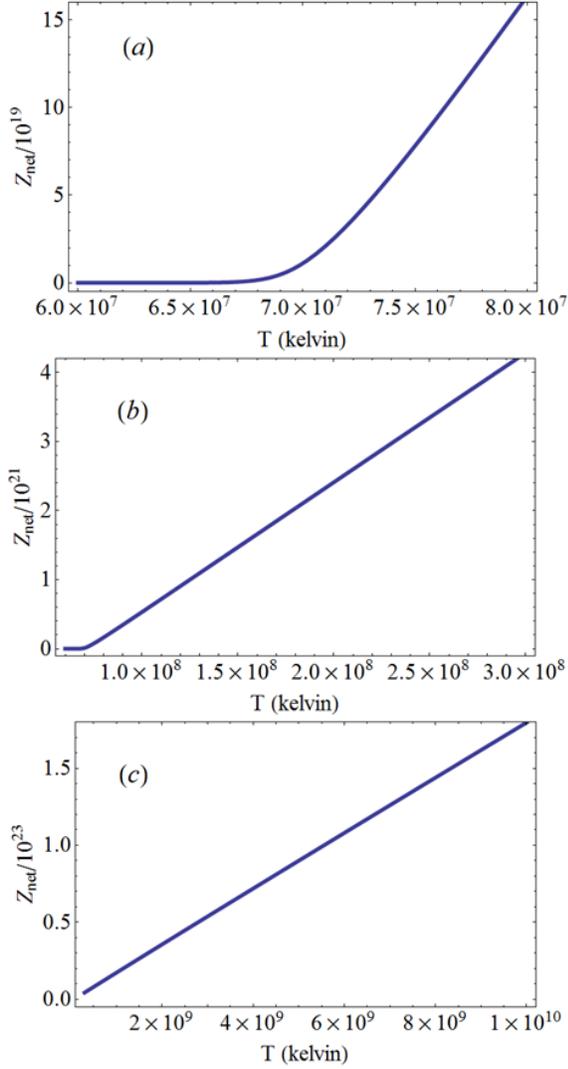

FIG. 5    Plots over a range of temperatures of the net charge vs. temperature in a stellar-mass core of radius 4x10⁶ m, due to electrons being excited into the 1ˢᵗ excited state of the extra-dimensional potential well.

As seen in Fig. 5a, the net charge begins to build up at a core temperature of just over 65 million Kelvin, however, the Coulomb energy term, $E_C$, in the exponent adequately damps the runaway build-up of net charge in the core, so that the dependence of $Z_{net}$ on $T$ becomes nearly linear, rather than exponential, as shown in Fig. 5b,c, over the range of temperatures expected in massive stellar cores.

The only way for an explosive charge imbalance to build up in the star due to electrons being excited into the extra dimensions is if the temperature of the core rapidly increases from the outside towards the center. Then the Coulomb energy term in the exponent of the Maxwell-Boltzmann fractional charge distribution would not apply at each spherical shell (since it only applies to the net charge *inside* the radius where the temperature is increasing). Another possibility is if the temperature increases in the center (r=0), causing an increase in $Z_{net}$ which then diffuses radially outward, which thus eliminates the Coulomb energy term, allowing a further increase in $Z_{net}$, and so on. Depending on the radial diffusion rate of positive ions, this process could lead to large Coulomb self-energy build-up due to a charge core. However, in a dense core, the radial diffusion rate is probably very slow.

A thorough application of this higher-dimensional theory to stellar evolutionary models is beyond the scope of this paper, but preliminary analysis indicates that physically improbable results are not evident. Further, this theory predicts an outflow of positive ions from stars which increases with core temperature, corresponding with observations of solar wind and mass loss from stars as they progress through post main-sequence stages [42].

The fraction of electrons excited into the 1st excited state of the higher-dimensional potential well will diminish if the temperature falls, but an interesting case is if the core becomes degenerate with a Fermi energy, $E_F > m_e c^2$. In this case, electrons may be blocked from dropping back into the ground state, since no unoccupied energy level exists for them to fall into. A degenerate core which is heating may have electrons excite into the higher-dimensions, but the fraction in the first excited level would be governed by Fermi-Dirac statistics rather than Maxwell-Boltzmann.

## I. Dark matter

Although we have postulated that particles in the 1st excited state of the higher-dimensional potential well quickly decay back to the ground state, it may be that primordial higher excited states of fundamental particles could be persistent from the time of the big bang until the present if zero-point energy does not sufficiently extend into the higher dimensions. If this is the case, even-n excited states (even values for both $n_\gamma$ and $n_\eta$) would manifest in our 3-d space with a non-zero probability. For example, referring to

Fig. 6, the plot of the probability distribution function $\left|\Psi_{2,2}(\gamma,\eta)\right|^2$ of the n=4 (n=$n_\gamma$+$n_\eta$, $n_\gamma$,$n_\eta$=2) excited

level shows a local maximum at $\gamma, \eta = 0$, and a probability of manifesting near 3-d space of about 3.95%

(given by integrating the normalized probability distribution function from $-\gamma_{\min}, -\eta_{\min}$ to $\gamma_{\min}, \eta_{\min}$,

where $\gamma_{\min}, \eta_{\min}$ are the values of the first zeros of $\left|\Psi_{2,2}(\gamma,\eta)\right|^2$ along the lines $\eta = 0, \gamma = 0$ ). In this

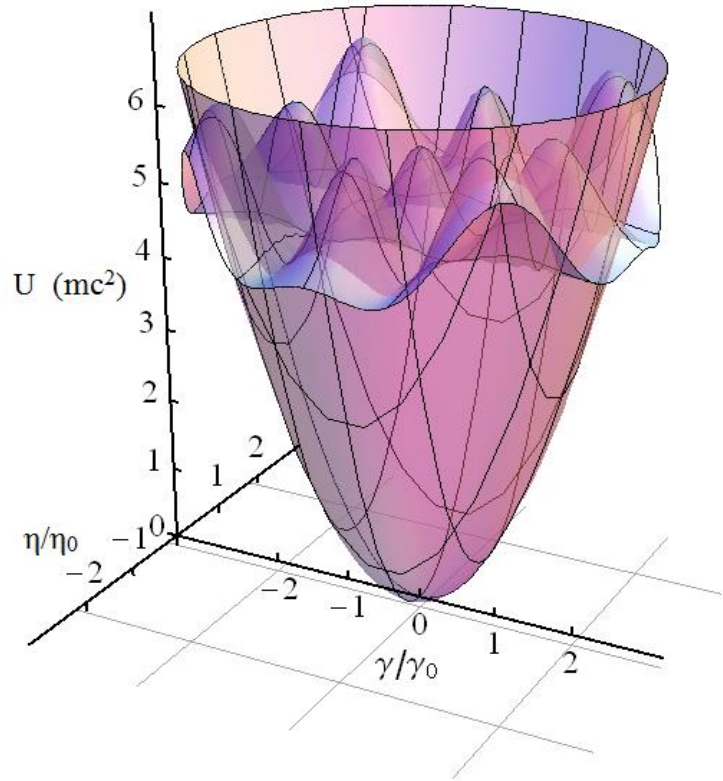

FIG. 6 (Color online) Depiction of the quadratic confining potential well in the higher dimensions $\gamma$ and

$\eta$. A symmetric excited state probability distribution (for $n_\gamma$=2, $n_\eta$=2) is superimposed, showing that in

such states, a probability maximum exists at $\gamma = \eta = 0$ (in 3-d space), even though most of the probability density resides in the higher dimensions.

model, I propose that dark matter consists of a ladder of these even-n excited states of the fundamental particles. The fractional probability of manifesting in 3-d space means that these excited states (dark matter) would only weakly interact with normal matter (ground states) and with electromagnetic radiation. The mass of these excited states, however is postulated to manifest in such a manner, that relative to the ground-state particle mass $m_0$, its mass is increased due to the sharper curvature of 3-d space (indicated by $\gamma_{min}, \eta_{min} < \gamma_0$), but diminished by the reduced probability for this state to manifest at $\gamma, \eta = 0$ in our 3-d space. The formula for the effective mass of a symmetric excited state is thereby given as,

$$m_{eff}/m_0 = \left(\frac{1}{W_0}\right)(\gamma_0/\bar{\gamma}_n)\int_{-\eta_{min}}^{\eta_{min}}\int_{-\gamma_{min}}^{\gamma_{min}}\left|\Psi_{n_\gamma, n_\eta}(\gamma, \eta)\right|^2 d\gamma d\eta . \qquad (18)$$

The pre-factor, $W_0$, in Eq. 18 has the value $W_0 = \int_{-\gamma_0}^{\gamma_0}\int_{-\gamma_0}^{\gamma_0}\left|\Psi_{0,0}(\gamma, \eta)\right|^2 d\gamma d\eta = 0.710145$, so that $m_{eff} = m_0$ for the ground state, n=0, and $\bar{\gamma}_n$ is the average of the first minima, $\gamma_{min}, \eta_{min}$, of the wavefunction $\left|\Psi_{n_\gamma, n_\eta}(\gamma, \eta)\right|^2$, for the state described by $n_\gamma, n_\eta$. For example, for the state $n_\gamma, n_\eta = 2$, we calculate $m_{eff}/m_0 = 0.07866$.

A plot of the net effective mass of the higher-dimensional excited states, summed over all the $n_\gamma, n_\eta$-even degenerate states for a given n=$n_\gamma$+ $n_\eta$, for n=2 to 100 is shown as $m_{eff}/m_0$ in Fig. 7 as the dashed curve (magnified by a factor of 10 to scale it up for viewing on the plot). The curves in Fig. 7 are power-law fits to the net and summed effective mass values calculated from the formula above ($m_{eff}/m_0$ was calculated explicitly for the first 90 excited states, corresponding to n=2 to n=24, with degeneracy 1+n/2) Also shown in Fig. 7 are the summed net effective mass values (solid curve) over the same range of n. This result provides a simple plausibility argument that dark matter can be accounted for by a finite number of symmetric higher-dimensional excited states (as shown in Fig. 7, $m_{TOT}/m_0 \approx 5.6$ for n=50, and $m_{TOT}/m_0 \approx 8.9$ for n=100). More detailed considerations involving superpositions of higher-order mixed modes of the excited states within the extra-dimensional confining potential would possibly modify this result.

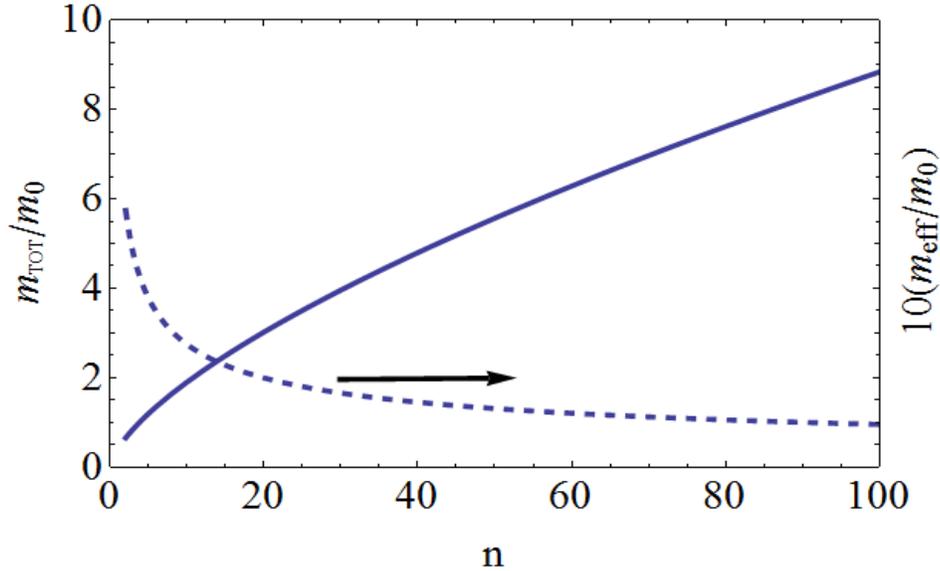

FIG. 7   Effective mass of the higher-dimensional excited states, summed over all the $n_\gamma$, $n_\eta$-even degenerate states for a given $n = n_\gamma + n_\eta$, (dashed curve), and the net sum of these effective masses for $n = 2$ to 100 (solid curve).

## 8.  CONCLUSION

A brief overview of the possible implications of the evanescent confinement of quantum particles in two extra dimensions is presented within a simplified model using non-relativistic quantum theory.  A harmonic oscillator 2-dimensional potential well (representing the $4^{th}$ and $5^{th}$ dimensions) is used to provide confinement to 3-d space.  Consistency with the Heisenberg Uncertainty Principle is used to establish the potential well parameters, giving a ground state energy equal to the rest-mass energy of the particle. The excited states of the higher-dimensional confining potential exist at energy levels separated by the rest-mass energy, suggesting the possibility that a particle may temporarily disappear into the higher dimensions if it absorbs an amount of energy equal to its rest-mass energy.  The theoretical framework presented is analyzed and shown to be consistent with a diverse variety of physical phenomena, including particle properties, special relativity, and astrophysics. It is also noteworthy that higher-dimensional considerations can empirically reproduce effects found in relativistic quantum mechanics, such as the *zitterbewegung* oscillatory motion of the particle, fermion spin, electron gyromagnetic ratio, and a velocity-dependent effective particle size. Further work is warranted, but it is hoped that the introduction of this concept will provide the stimulus for fruitful ongoing analysis.